# Bird Species Classification And Acoustic Features Selection Based on Distributed Neural Network with Two Stage Windowing of Short-Term Features

Nahian Ibn Hasan[a],[*]

[a]*Department of Electrical and Electronic Engineering, Bangladesh University of Engineering and Technology, Dhaka, Bangladesh*

**Abstract**

Identification of bird species from audio records is one of the challenging tasks due to the existence of multiple species in the same recording, noise in the background, and long-term recording. Besides, choosing a proper acoustic feature from audio recording for bird species classification is another problem. In this paper, a hybrid method is represented comprising both traditional signal processing and a deep learning-based approach to classify bird species from audio recordings of diverse sources and types. Besides, a detailed study with 34 different features helps to select the proper feature set for classification and analysis in real-time applications. Moreover, the proposed deep neural network uses both acoustic and temporal feature learning. The proposed method starts with detecting voice activity from the raw signal, followed by extracting short-term features from the processed recording using 50 ms (with 25ms overlapping) time windows. Later, the short-term-features are reshaped using second stage (non-overlapping) windowing to be trained through a distributed 2D Convolutional Neural Network (CNN) that forwards the output features to a Long and Short

---

[*]Corresponding author.
  *Email address:* nahianhasan1994@gmail.com (Nahian Ibn Hasan)

Term Memory (LSTM) Network. Then a final dense layer classifies the bird species. For the 10 class classifier, the highest accuracy achieved was 90.45% for a feature set consisting of 13 Mel Frequency Cepstral Coefficients (MFCCs) and 12 Chroma Vectors. The corresponding specificity and AUC scores are 98.94% and 94.09%, respectively.

## 1. Introduction

Birds are one of nature's most divine creatures that have a significant collaboration in maintaining the ecological balance throughout the world. There are over 10,000 bird species in the world. Unfortunately, due to poaching and other artificial or natural imbalance, many of these significant species are on the verge of extinction. As a result, many unwanted effects are occurring in nature due to the reduction in the number of birds. For example, birds are placed very high in the food chain. Hence, they integrate the changes that occur at lower levels. As a result, birds are good indicators for ecological changes or the devastating environment in the lower levels. That is why it is so important to know about different species of birds in an area as well as their number for having a comprehensive knowledge of the local environment. However, birds are often heard rather than seen by us. That is why it is an excellent area of practice to identify birds by analyzing their calls and sounds. There have been many endeavors for identifying and classifying birds from their natural habitat, but there are many sources of noises such as environmental, other living things, and different bird species sound in the background. Hence, most of the time, the recordings are classified by domain experts, which is a very time-consuming and hectic task. Therefore, automatic detection of birds is a dire need for efficient and fast identification. Traditional audio signal processing techniques and machine learning-based techniques have started playing a significant role in this respect. But using only either of these approaches face some challenging situations since many of the databases have domain mismatches between the training and test recordings. For that reason, the classification and identification accuracy are very low. Hence, a comprehensive algorithm is necessary, which can utilize the benefit of both of these approaches simultaneously.

Among the reported results, a bird species classification scheme has been suggested by [1] that was addressed using spectrogram textural features and the dissimilarity framework. According to them, this reduces the sensitivity of the model to the increase of the number of classes. Besides, Kahl et al. [2] introduce a convolutional neural network (CNN) for large-scale bird song classification. They trained a CNN with the spectrograms from bird sounds. On the other hand, Koh et al. [3] did the same thing as in [2] with the exception that they have trained the data with predefined CNN structures like 'Resnet'[4] and 'Inception'[5]. Similarly, Incze et al. [6] also trained a pre-trained neural network, named 'MobileNet'[7] with the spectrograms obtained from the birdcalls.

However, Xie et al. [8] has used both visual and acoustic features to train a CNN and combine the results of both domains to classify a bird sound. They have transformed the audio data through Constant Q-Transform (CQT), which is the input feature to CNN. For acoustic features, they have chosen spectral centroid, spectral bandwidth, spectral contrast, spectral flatness, spectral rolloff, zero-crossing rate, the energy of the signal, and Mel Frequency Cepstral Coefficients (MFCC). For acoustic and visual feature classification, they have compared the results of K-NN and Random -Forest classifier. Also, they have analyzed the birds' sound through CNN networks. Likewise, Bold et al. [9] has also employed both audio and visual data to train a CNN and then used fusion strategies to combine these cross-domain. However, such an approach, which requires both audio and image of the bird, poses some challenges in practical application. Most of the time, in nature, birds are heard rather than seen. Hence, automatic detection from the only audio sound is the most feasible approach but still a challenging one.

At the same time, Lasseck et al. [10] has proposed that training a fine-tuned pre-trained deep CNN model with Mel-spectrogram provides a good performance in terms of Area Under Curve (AUC) evaluation metric. Apart from all of these, according to Briggs et al. [11], the probabilistic model of short-term feature window of the audio recording and then applying Bayes Risk Minimizing Classifier provides a significant improvement to the classifier. In this case, they have used different acoustic features per frame, such as MFCCs, spectrum bandwidth, and spectrum density. They have also reported that such an approach provides similar accuracy to an SVM.

A little comprehensive real-time classifier approach was adopted by Raghuram et al. [12]. They have extracted audio frame features, such as harmonic product spectrum, energy, auto-correlative features, Tempo, pulse clarity, MFCCs, etc. Then they trained different machine learning methods like Naive Bayesian approach, SVM, Random Forest, and Neural Network. They also analyzed the classification accuracy between bird songs and bird calls. Later they combined the birds' habitat feature, bird call-type, audio recording features, and bird weight predictor to predict the bird species. Mel spectrogram was also used by Schluter et al. [13] as an input feature to the ensemble CNN and Multilayer Perceptron Models (MLPs) along with other secondary features like bird geographic location. Apart from the acoustic frequency-based spectrograms, different types of spectrograms like Gabor Transformed spectrograms have been used as the primary input feature for CNN by Heuer et al. [14]. They also analyzed the effect of different window lengths in classification performance.

One of the main purpose of this paper is to analyze the effect of acoustic feature selection in deep-learning based bird-call classification task. The aim is not to achieve the best accuracy, rather provide a comprehensive idea about feature selection. This paper analyzes different short-term window acoustic features as primary features. Later these features are provided as input to the distributed CNN network followed by an LSTM network to classify the bird species. The study only incorporated bird recording irrespective of bird call type, geographic location, recording type, noise type, etc. Hence, the proposed method provides a general scheme of birds call classification using a CNN-LSTM based network.

The network size is kept small for real-time applications. Besides, this paper represents a comprehensive study with different combinations of a variety of acoustic features. Moreover, adaptive thresholding, for voice activity detection, helps to detect the activity of the foreground voice in the recordings. Hence, the proposed approach is designed by considering the real-time application and general classification scheme irrespective of other secondary features that may vary from geography, environment, recording tool.

In the following section (section 2), a brief overview of the proposed method is presented. Section 3 represents short discussions on each of the acoustic features. Next, the extracted features (of varying length) are segmented into equal lengths and reshaped to pass through the distributed CNN. However, the segmented features are passed in parallel to the distributed CNN. The modified segmented features, from the output of CNN, flow to the Long and Short Term Memory (LSTM) network. The resultant feature matrix of the LSTM layer flows through two densely connected layers. The final dense layer classifies the bird-call/song into respective classes.

## 2. Proposed Method

Fig. 1 illustrates the proposed method as a block diagram. The raw signal is processed through adaptive thresholding for foreground voice activity detection in the record. Next, the signal is analyzed to gather short-term features based on short-term overlapping windows. The short-term features include zero-crossing rate (ZCR), energy, the entropy of energy, spectral centroid, spectral spread, spectral entropy, spectral flux, spectral roll-off, 13 Mel frequency cepstral coefficients (MFCCs), 12 chroma vectors, and chroma deviation.

## 3. Short Term Features

This section briefly describes the the short term features that have been investigated throughout the study. Apart form the theoretical discussion, there

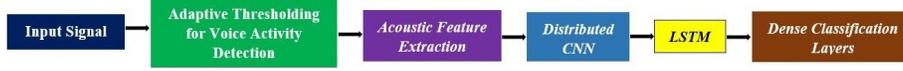

Figure 1: The proposed method for classifying bird-calls/sounds. First the voice activity in the long record is extracted through adaptive threshold technique. Then different acoustic features from the signal are passed through a distributed CNN network, followed by an LSTM network. The final densely connected layer classifies the signal.

are subsections where the application of the theory is illustrated based on the dataset.

### 3.1. Zero Crossing Rate

The zero-crossing rate of a signal denotes the number of times the signal changes its sign along the zero axis line in a time frame. The zero-crossing rate is useful for detecting the presence of voice activity in a signal. It also describes the smoothness of the signal within the time frame. The zero-crossing rate can be expressed with the following formula-

$$\text{Zero-crossing-rate} = \frac{1}{L-1} \sum_{x=1}^{L-1} I.W_x W_{x-1} \qquad (1)$$

Here, W=signal, $W_x = x^{th}$ sample of the signal, **x=index for the signal samples**, L=length of the signal, I=indicator function.

### 3.2. Energy of the Signal

Signal energy is defined as the sum of squares of the signal sample values. Then it is normalized with respect to the length of the time frame. It is formulated as-

$$\text{Normalized signal energy} = \frac{1}{L} \sum_{n=n_1}^{n_2} |x(n)|^2 \qquad (2)$$

here, $(n_1, n_2)$ = sample range in the time frame, x=signal, L=length of the time frame.

### 3.3. Entropy of Energy (EE)

The entropy of energy (EE) defines the distribution of energy over a time frame. Hence, first, the energy of the signal is calculated according to equation 2. Then the entropy of that signal is calculated. The following formula expresses the entropy of the signal as-

$$\text{Entropy of Energy} = -\frac{\sum_{j=1}^{M} E(j) log_2 E(j)}{log_2 M} \quad (3)$$

here, E(j)=the energy of the signal within a time frame. M=length of the energy signal.

### 3.4. Spectral Centroid

Spectral centroid defines the center of gravity of the spectrum of a signal within the time frame. Spectral centroid of a sound is mostly associated with the brightness of the sound. Spectral centroid of a signal can be formulated as follows according to [15]-

$$\text{Spectral centroid} = \frac{\sum_{k=b_1}^{b_2} f_k s_k}{\sum_{k=b_1}^{b_2} s_k} \quad (4)$$

Here, $f_k$ = frequency in Hz corresponding to bin k, $s_k$ = the spectral value at bin k, b1,b2 = band edges, in bins, over which to calculate the spectral centroid.

### 3.5. Spectral Spread

The spectral spread of a signal describes the average deviation of the spectrum around its centroid. It is considered to be associated with the bandwidth of the signal. The signal with different kinds of noise or noise-like features constitutes a large spread, while individual signals without background noise (like a human voice, birds call) and with specific tones have a narrow spread. Usually, the spectral spread is expressed in the range of zero to one by dividing it with the highest center frequency. Sometimes this feature is called the second central moment of the spectrum. However, the spectral spread can be expressed as follows according to [15]-

$$\text{Spectral spread} = \sqrt{\frac{\sum_{k=b_1}^{b_2}(f_k - \mu_i)^2 s_k}{\sum_{k=b_1}^{b_2} s_k}} \quad (5)$$

here, $f_k$ = frequency of the $k^{th}$ bin in hertz (Hz), $s_k$ = spectral value at $k^{th}$ bin, $(b_1, b_2)$ = spectral range of the signal over which to calculate the spectral spread (In our case, it should be the short-term feature window), $\mu_k$ = spectral centroid.

*3.6. Spectral Entropy (SE)*

Spectral entropy of a signal signifies its spectral power distribution. According to spectral entropy concept, any signal's normalized power is considered to be a probability distribution and then Shanon entropy is calculated, which is called signal entropy. Spectral entropy is frequently used for speech analysis and recognition [16].

Say, for a signal W(n), the power spectrum is $Q(m) = |W(m)|^2$, where W(m) is the discrete Fourier transform of W(n). The probability distribution P(m) can be expressed as-

$$P(m) = \frac{Q(m)}{\sum_j Q_j} \quad (6)$$

Next, the spectral entropy is calculated as follows-

$$SE = -\sum_{j=1}^{M} P(m) log_2 P(m) \quad (7)$$

If we normalize the spectral entropy-

$$SE_n = -\frac{\sum_{j=1}^{M} P(m) log_2 P(m)}{log_2 M} \quad (8)$$

Here, M=total number of frequency points in the spectrum. The term, $log_2 M$ denotes the maximal spectral entropy of white noise, which is uniformly distributed in the overall frequency domain.

*3.7. Spectral Rolloff*

Spectral roll-off defines the specific frequency (Hz) below which a specific percentage of the spectral energy resides. The range of this threshold frequency lies within (in general) 85% to 95%. This feature identifies the silence in any voice signal. In our case, we considered this threshold to be 90% for feature analysis and preparation for the neural network. We calculated the spectral roll-off frequency in each window frame of the signal. Spectral roll-off is calculated using the formula described in [17] as follows. For the roll-off frequency of $f_i$-

$$\sum_{k=b_1}^{f_i} s_k = R \sum_{k=b_1}^{b_2} s_k \tag{9}$$

Here, R = Threshold percentage (in our case it is 90%), $(b_1, b_2)$ = spectral range, $s_k = k^{th}$ spectral value.

*3.8. Spectral Flux*

Spectral flux describes how frequently the power spectrum of a voice/signal changes over the time. In our case, we calculated the spectral flux by taking the squared difference between the normalized magnitudes of the spectra of the two successive frames. Spectral flux is first described in [17]. Spectral flux is very useful for the classification of speech sounds. Spectral flux can be mathematically defined in the following manner according to [17]-

$$\text{Spectral flux (t)} = [\sum_{k=b_1}^{b_2} |s_k(t) - s_k(t-1)|^P]^{\frac{1}{P}} \tag{10}$$

Here, $s_k$ is the $k^{th}$ spectral value, $(b_1, b_2)$ = spectral range of the spectrum, P=norm type (in our case P=2).

*3.9. Mel Frequency Cepstral Coefficients (MFCC)*

Mel Frequency Cepstrum (MFC) defines the short-term power spectrum of a sound. It is dependent on the cosine transformation of the logarithmic power spectrum based on the non-linear Mel scale of a frequency. The shape of the human vocal tract (consisting of mouth, teeth, tongue, larynx, etc.) define the generated voice. The accurate sound and phoneme can be determined if the shape of the vocal tract is accurately known or modeled. The envelope of the short-term power spectrum of the sound represents itself as the shape of the vocal tract. Mel Frequency Cepstral Coefficients (MFCC) models this envelope. The same idea is applied to determine the MFCC of a bird song or sound. The different structures of the vocal tract of birds (also known as syrinx) are represented in [18]. The chart of Fig. 2 [19] illustrates the general algorithm of MFCC.

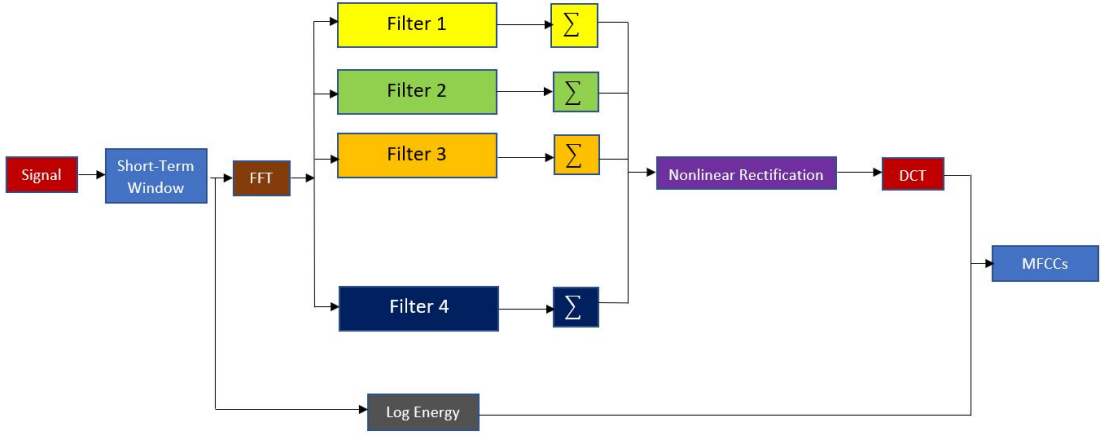

Figure 2: Algorithm for determining mel frequency cepstral coefficients (MFCCs). The Fourier transformed signal of the record is passed through a Mel filter bank. The Mel-scale has a logarithmic spacing of over 1 kHz. 40 Mel-filters have been applied. Next, Discrete Cosine Transform (DCT) of log transformed signal of each of the output from the filter banks provides 40 different DCT coefficients. The first 13 of these coefficients are the Mel-13 frequencies, which have been used as features in this paper. A detailed explanation of the algorithm can be found in [19] and [20]

here, FFT denotes the 'Fast Fourier Transformation', DCT stands for 'Discrete Cosine Transformation'. There are 13 coefficients extracted for a record. The signal is first decomposed into short term windows. In our case, the length of the window is 50ms with an overlap of 25ms. Say, for the $j^{th}$ frame of the signal, the discrete Fourier transform (DCT) of the signal is $W_j(k)$ -

$$W_j(k) = \sum_{x=1}^{M} w_j(x)h(x)exp(-i2\pi kx/M); \quad [1 < k < Q] \quad (11)$$

here, $w_j(x)$ denotes the time domain signal where $(1 < x < M)$, M=total number of samples in the frame, h(x) is a hamming window, Q=length of the DFT. The periodogram based power spectral estimate of the $j^{th}$ frame can be expressed as follows-

$$Y_j(k) = \frac{1}{M}|W_j(k)|^2 \qquad (12)$$

Next, we apply 40 MFCC-filters (27 logarithmic filters and 13 linear filters) to the periodogram based power spectral estimate ($Y_j$). This results in 40 different numbers that represent the energy in each of those filters. After that, taking the logarithm of each energy leaves us with 40 different log energy levels. Next, we take the discrete cosine transformation (DCT) of the 40 log energy levels to gain 40 DCT coefficients which can be formulated as follows-

$$G(x) = \left(\frac{2}{S}\right)^{\frac{1}{2}} \sum_{u=0}^{S-1} F(u) cos\left[\frac{\pi x}{2S}(2u+1)\right] f(u) \qquad (13)$$

Here, G(x)=DCT coefficients, S=length of the input signal (which is 40), u=variable for 40 energy levels, x ranges from 1 to maximum number of DCT coefficients (in our case, we want 40 coefficients), f(u) = log energy levels of the 40 MFCC filter banks, F(u) is denoted as follows-

$$F(u) = \begin{cases} \frac{1}{\sqrt{2}} & ; u = 0 \\ 1 & ; otherwise \end{cases}$$

Finally, we keep the first 13 DCT coefficients for the analysis. These 13 coefficients are the MFCC of a single frame. Fig. 3 represents the MFCC features of a single class. Most often, MFCC is called static feature aggregation method as it is applied on a single frame. They have been used by most scholars for frame-wise speech detection and music modeling [21][22][23][24].

*3.10. Chroma Vector Features*

Musical audio analysis frequently involves chroma features. In chroma analysis, the method projects the whole spectrum of the frame onto 12 bins that represent the 12 different pitch classes or semitones or chroma.

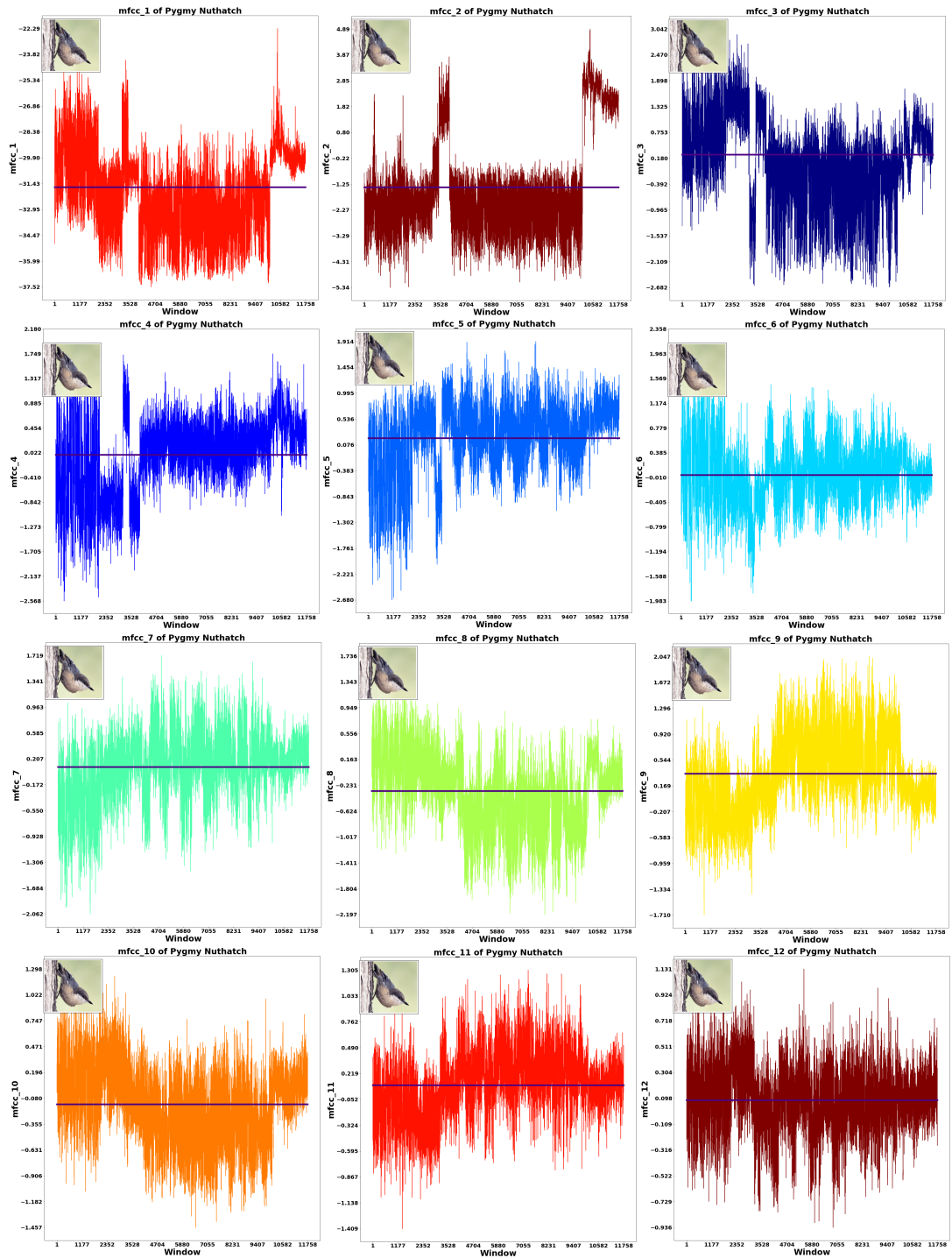

Figure 3: Mel frequency cepstral coefficient (MFCC) variation of a single recording. The variations are shown for active foreground voice (after voice activity detection using adaptive thresholding). The black horizontal line resembles the mean value of the variation.

In the musical analysis, notes which are exactly one octave apart are similar in nature. Let's say we don't know the absolute frequency (original octave). But, still, a significant amount of musical information about the audio signal can be achieved by the chroma distribution. Apart from this, some musical similarities are easily recognizable, which are not usually apparent in the original spectrum of the audio. This kind of feature can help classify songbirds or other birds whose tone has certain musical characteristics. Chroma features are sensitive to variations in timbre. They almost accurately correlate to the musical harmony. Hence, the chroma feature is often used by many scholars for analyzing music or musical tones [25][26][27][28][29]. That is why using chroma features in birds call/song classification could be a good choice. The 12 chroma vectors are named as 'A', 'A#', 'B', 'C', 'C#', 'D', 'D#', 'E', 'F', 'F#', 'G', 'G#'. Let's say we have a time window (or frame) of a signal. A given chroma feature provides a single coefficient that represents all the information within that frame. The shifting of the window along the signal generates a sequence of chroma features. Each of these features is a way to express how the pitch content of that frame is inherent in the 12 chroma vectors. Hence, we get a time-domain image of the chroma vectors, also known as 'Chromagram'. Short term Fourier transformations are used for calculating chroma vectors. A detailed representation can be found in [30][31]. Fig. 4 represents several chroma vectors of a single recording. Also, Fig. 5 illustrates chromagrams of different classes.

*3.11. Chroma Deviation*

Chroma deviation denotes the standard deviation of the 12 chroma vectors. It can be formulated as follows-

$$\text{Standard deviation}(\sigma) = \sqrt{\frac{\sum (C_i - \mu)^2}{12}} \qquad (14)$$

Here, $C_i = i^{th}$ chroma vector, $\mu$ = mean of the chroma vectors. This chroma deviation quantifies the inter-variability of the chroma vectors, whether there is any existence of multiple pitches in the same time frame.

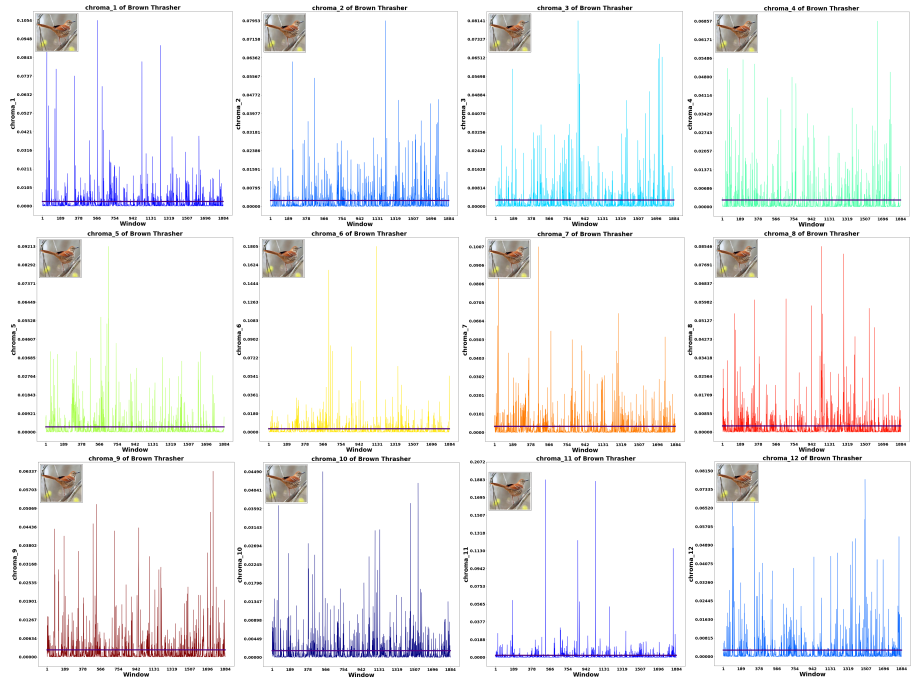

Figure 4: 12 Chroma-Vectors variation of a single recording. The chroma vectors are shown for active foreground voice (after voice activity detection using adaptive thresholding). The black horizontal line resembles the mean value of the variation.

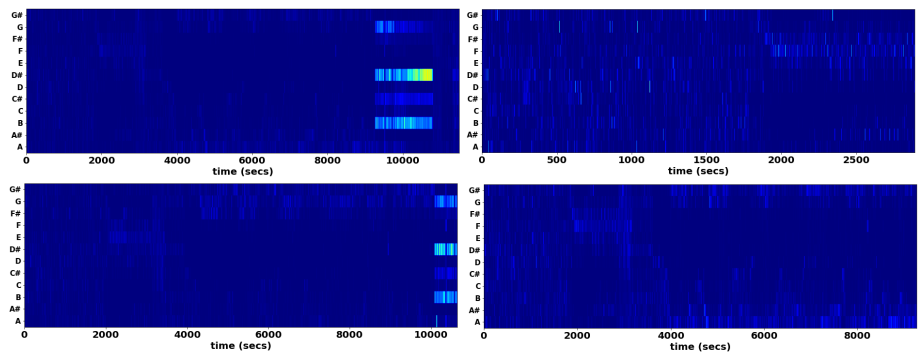

Figure 5: Chroma-Vectors variation as chromagram for varying classes. The chromagrams are shown for active foreground voice (after voice activity detection using adaptive thresholding).

## 4. Training Methodology

In this section, a detailed discussion of the training procedure is shown. First, there is a brief discussion on the database, followed by the distributed CNN-LSTM architecture, feature learning procedure, hyper-parameters and non-hyper-parameters of training, and evaluation metrics for measuring the network performance.

*4.1. Database*

The database has been collected from Cornell Lab of Ornithology's Center for Conservation Bioacoustics (CCB) [32][33]. For analysis purposes, data of 10 different bird species are collected, illustrated in Fig. 6 mentioning their species name and scientific name in parenthesis. The images in Fig. 6 are collected from Cornell ebird data repository [34]. The recordings are a mixture of a single channel (mono) or double channel (stereo) signals with different sampling rates. Most of the signals are of 44100Hz and 48000 Hz). The recordings are from different geographic locations. The sounds belong to normal bird-call, bird-song, complex interaction calls, wee-oo or zwee-oo call, flight call, alarm call, whisper song, threat calls, juvenile calls, begging call, juvenile babble, etc. with some recordings having background noise like airplane sounds, other animals calls, etc. Most of the audio recordings belong to 128000 bit-per-seconds (bps) and 320000 bps. There are 921 records of bird-songs/calls, among which 737 records belong to the training set and 184 belong to the test set. Following the standard rule of thumb, the dataset has been divided in a [80-20]% fashion into train and test data. Fig. 7 illustrates the data distribution of each class. Fig. 7(a) denotes the number of train data for different classes, whereas Fig. 7(b) illustrates the number of test data for different classes.

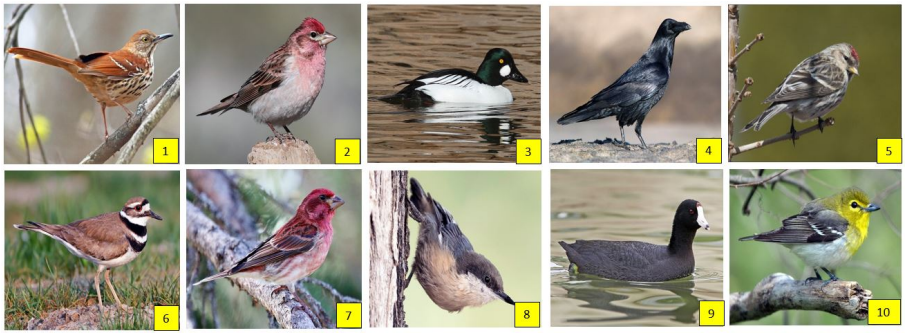

Figure 6: Some sample bird species from the database. [1]Brown Thrasher (*Toxostoma rufum*), [2]Cassin's Finch (*Haemorhous cassinii*), [3]Common Goldeneye (*Bucephala clangula*), [4]Northern Common Raven (*Corvus corax*), [5]Common Redpoll (*Acanthis flammea*), [6]Killdeer (*Charadrius vociferus*), [7]Purple Finch (*Haemorhous purpureus*), [8]Pygmy Nuthatch (*Sitta pygmaea*), [9]American Coot (*Fulica americana*), [10]Yellow-throated Vireo (*Vireo flavifrons*).

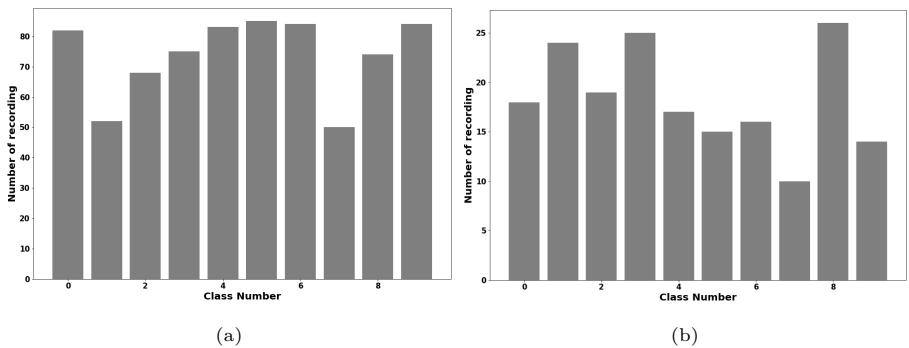

(a) (b)

Figure 7: Class specific data distribution in the database. (a) Train data distribution in the dataset for 10 classes. (b)Test data distribution in the dataset for 10 classes.

## 4.2. Pre-processing

Most of the recording in the database has background noise, which includes airplane sounds, other animal calls, other bird-calls, etc. Besides, a record has multiple segments that lack any foreground sounds of interest. Hence, a simple and common preprocessing task is necessary to remove the silent portions of the recording. In this respect, an adaptive threshold technique is selected for foreground voice activity detection in the recording. The voice activity detection would not have been necessary if there was a separate background class along with the 10 bird-classes. But training that background class will become a challenge as it will highly depend on the data source, and the result obtained from such study will pose ambiguity. Hence, to keep the generalized applicability of the proposed method, it is necessary to perform voice activation.

Foreground sound denotes the sound of interest (like the specific bird-song, bird calls, etc. that is to be recognized).It is assumed that the background noise will always be there throughout the recording. Whenever there is any occurrence of foreground voice (bird call/song), the signal strength will be higher. Therefore, we have used adaptive thresholding for foreground voice activity detection. The silent portion of the recording signifies the time-frame of the recording where the foreground sound is absent. The active part of the recording denotes the segments of the audio recording where the foreground sound is present. Therefore, in this preprocessing technique, the purpose is to remove the silent portion from the signal. However, differentiating foreground and background sounds in the active part of the record is challenging. Hence, it is expected that the neural network can differentiate the background noise in the active part of the record. The purpose of the neural network is to learn prominent features of different bird's voice sounds and differentiate among them in the presence of background noise.

The adaptive threshold technique uses dynamically changing threshold value for the voice strength based on short term windows. For now, an initial threshold value of 0.6, short term window of 50ms and window shift of 25ms are used. For the $i^{th}$ window, the signal is denoted as $X_i$. For applying threshold, the signal is first converted into one sided signal by squaring the sample amplitude values as follows-

$$X_i = X_i^2 \tag{15}$$

Next, the primary threshold value is calculated based on the features of $i^{th}$ frame as follows-

$$T_i = X_{i,min} + (X_{i,pp} * T) \tag{16}$$

here, $T_i$ = primary threshold value for the $i^{th}$ frame, T = global threshold value applied to any frame, $X_{i,min}$ = lowest amplitude of the signal in the $i^{th}$ frame, $X_{i,pp}$ = peak to peak signal value in the $i^{th}$ frame.

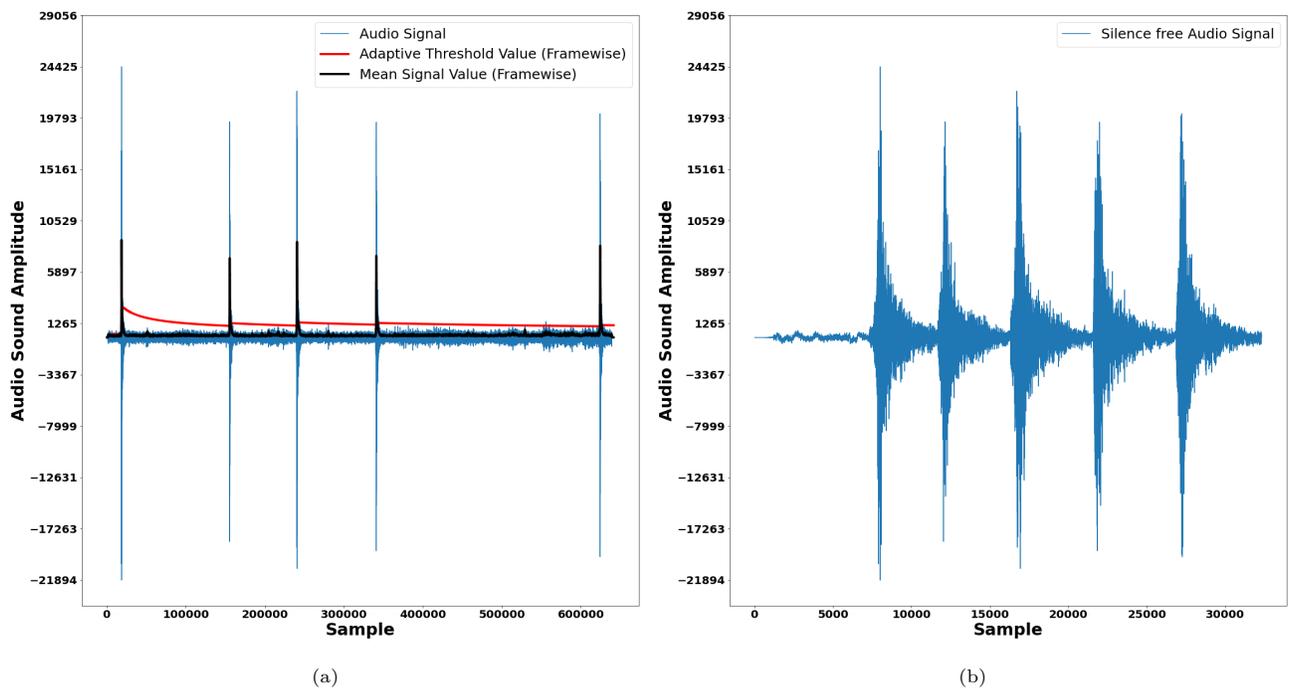

Figure 8: (a) Adaptive threshold technique for voice activity detection in a bird-call. Whenever the mean value of the signal in a frame is lower than the adaptive threshold value for that frame, the frame is considered to be a silent frame. (b) Resultant signal after removing silent portions of the record.

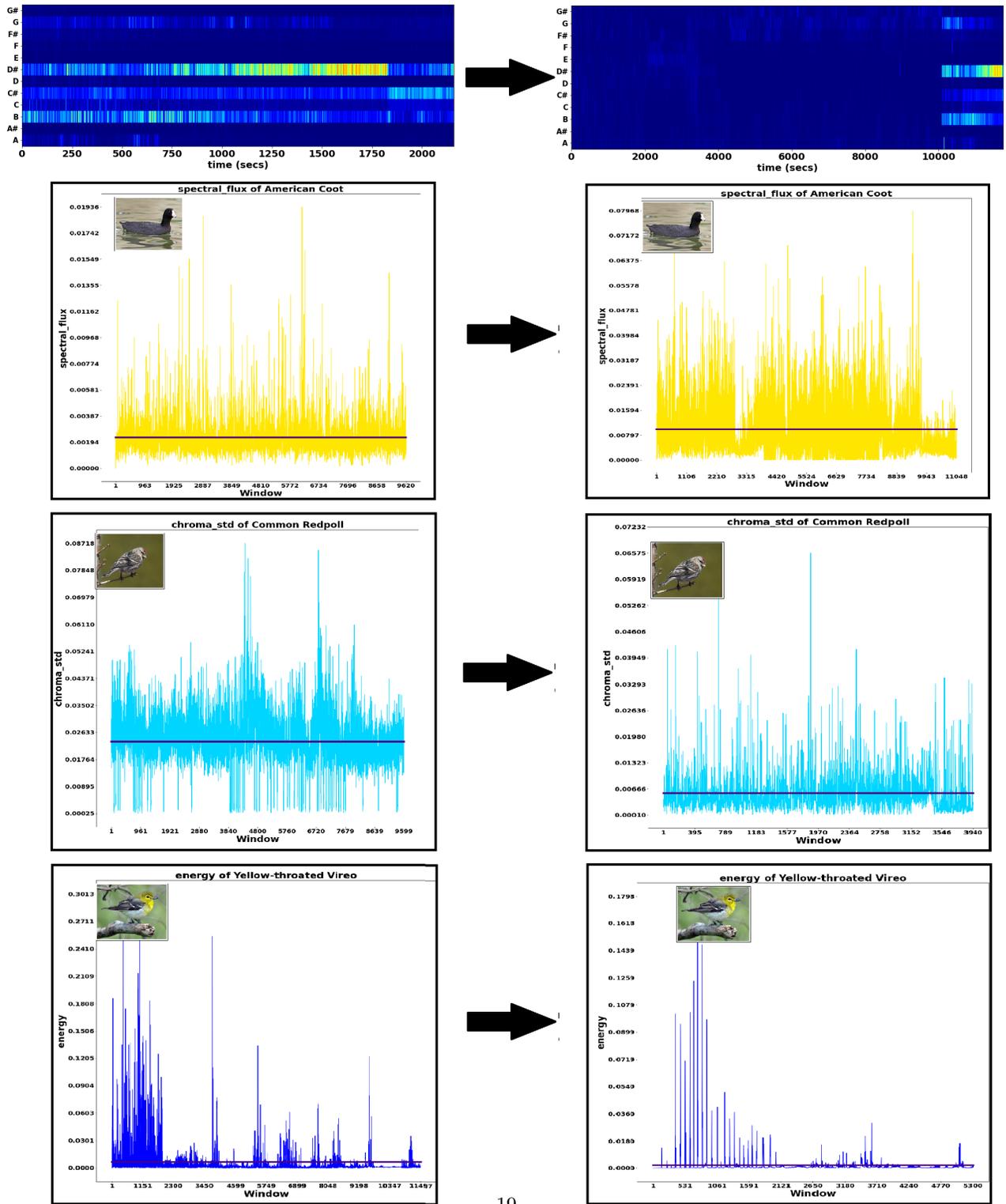

Figure 9: Voice-activity-detection effect on different acoustic features. The left hand side images are the features of the raw signal, the right hand side images denote the same feature of the same recording for the foreground active voice part only. The foreground voice activity has been detected through adaptive thresholding strategy.

Finally, the adaptive threshold value for the $i^{th}$ frame is calculated as follows-

$$T_{i,ad} = \frac{(i-1)*T_{i-1,ad} + T_i}{i} \qquad (17)$$

here, $T_{i,ad}$ = adaptive threshold value for the $i^{th}$ frame, $T_{i-1,ad}$ = adaptive threshold value for the $(i-1^{th})$ frame. Hence, the adaptive threshold value for any frame is calculated from the primary threshold value of the frame and the adaptive threshold value from its previous frame. The index i starts from 1 (i.e. first frame) to onward. Hence, for equation 17, a starting value of $T_{0,ad}$ is necessary which is assumed to be zero. Besides, for applying equation 16, a global threshold value of T is needed to be defined. For the purpose of this paper, T is set to be 0.6.

Finally, if the signal's mean value $X_{i,mean}$ in a frame is lower than the adaptive threshold of that frame ($T_{i,ad}$), then the frame is considered to be a silent frame (for example, no foreground voice activity). Sakhnov et al. [35] reports the general algorithm for foreground voice activity detection. Fig. 8 illustrates a sample bird-call record with an adaptive threshold. Fig. 8(a) represents the feature of the original image, whereas Fig. 8(b) represents the corresponding feature for the active foreground voice part after voice activity detection. The effect of voice activity detection is illustrated in Fig. 9.

*4.3. Neural Network Architecture*

Fig. 10 depicts the neural network used for training the audio signals. This is a detailed version of Fig. 1. In Fig. 10, individual colors in the features segment represent individual short-term features. Hence, the same color blocks illustrate the passage of a single feature through the network. The network weights update after each cycle of segmented features. Each cycle constitutes a bunch of segmented features, each of which belongs to a different mother feature. Hence, a single cycle constitutes segmented features of all the short-term features.

Since the signals are of varying length, so are the features. The features need to be segmented into equal lengths before passing them to the distributed

CNN. The CNNs are mostly used on images. The convolution operation on an image extracts salient features of the image. However, in our case, we have a 1D signal. The idea is - forming a 2D image from a 1D vector and applying the convolution operation on that image. It is expected to extract the salient features of that image, which corresponds to the original vector. In this case, the short-term features are segmented into chunks of 1000 samples. That means, the method is applying a second stage windowing without overlapping. Each of these segments is a single feature, which is derived from the mother feature vector. This is repeated for all of the short-term features. All the segmented parts are passed one-by-one through the distributed CNN network.

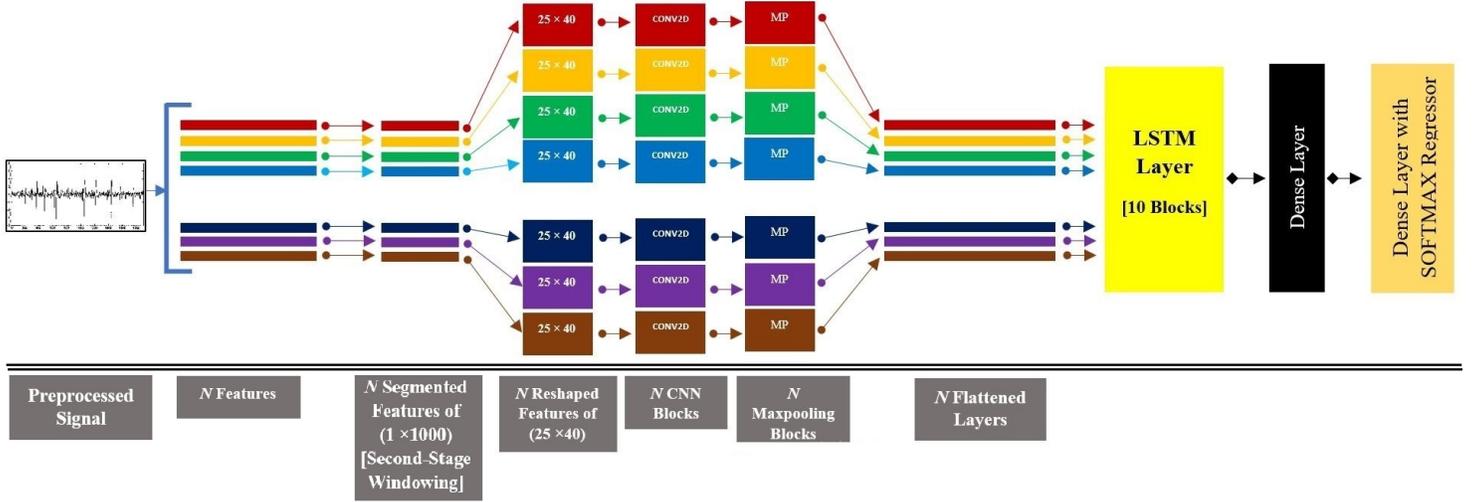

Figure 10: Distributed Convolutional Neural Network (CNN) with Long and Short Term Memory (LSTM) network. The input signal denotes the preprocessed signal. There are **N** features extracted from the preprocessed signal. Each of the features are divided into non-overlapping segments of size $(1 \times 1000)$. This is second stage windowing. Next, distributed convolutional operations along with maximal pooling operations are applied separately on each feature segment. Finally, the outputs from the distributed CNN networks are passed to **LSTM** layer containing 10 blocks. Finally, two **Dense** layers are applied for classification task.

The distributed CNN network applies a separate CNN network to individual feature, independently. But before applying a 2D convolutional operation, the segmented feature of size $(1 \times 1000)$ is reshaped into a single channel image of size $(25 \times 40)$. Next, the network applies 2D convolutional operations to the images derived from segmented features of different short-term features. Each convolutional operation constitute 50 kernels of size $(2 \times 2)$ with a stride of $(1 \times 1)$. No padding is used for the images. Each of the kernel weights are initialized with Xavier uniform initializer [36]. The bias vectors are initialized with zeros. No regularization function is applied to any of the kernel weights or bias vectors. The output of the convolutional layer is then passed through a Rectified Linear Unit (ReLU) activation function [37]. No activity regularizer is applied here. No constraint function is applied to either kernel weights or bias vectors. The reason for applying no regularization/ constraint is because the problem is well defined. Also, there are many preprocessing steps before applying the CNN. Usually, the regularization helps for a ill-posed problem.

The output from the convolutional block is passed through a down-sampling operation of Max-Pooling. The 2D Max-pooling is performed on the convolutional feature with a window size of $(2 \times 2)$ and a stride of $(2 \times 2)$. Although it is customary to pad with zeros in image-based analysis (i.e. image classification, object detection and recognition), it is not applied here due to the context.

Next, the reduced convolutional feature images are flattened to convert back into 1D features. Hence, the distributed CNN network applies convolutional and max-pooling operations on the 2-dimensional segmented features separately and generated modified segmented features at the end. If there are N features primarily, there will be N modified-segmented features.

The Long and Short Term Memory (LSTM) networks are mostly used for time-varying signals so that the local and global features of timed-signal can be learnt efficiently. Since, in our case, we have a time-varying signal as well as a silence-removed-signal, LSTM is supposed to be the best tool for feature extraction. The detailed explanation of LSTMs can be found in [38]. The LSTM layer consists of 10 LSTM blocks. The LSTM layer has hyperbolic tangent or tanh activation function and sigmoid as recurrent activation function. The LSTM kernel weights are initialized with Xavier uniform initializer [36] like the distributed CNN network. The initializer for the recurrent kernel weight matrix is orthogonal, which is used for the linear transformation of the recurrent state.

The bias vectors are initialized to zeros as well like distributed CNN. According to Jozefowicz et al. [39], 1 is added to the bias of the forget gate of the LSTM block at initialization. None of the kernel regularizer, bias regularizer, activity regularizer, and recurrent regularizer is used in the LSTM block. Also, there is no kernel constraint, bias constraint, or recurrent constraint. LSTM block does not contain any dropout function. The LSTM layer experiences an input tensor of shape (number of features $\times$ 1000). The output of the LSTM layer is passed to a densely connected layer.

The first dense layer consists of 10 neural nodes. There is no activation function applied to the first Dense layer as the following dense layer finally performs the classification by softmax regressor. It is assumed that at the end of the network, adding an activation function extracts no important features of a signal. The kernel weights in both of the Dense layers are initialized with Xavier uniform initializer [36] and the bias vectors are initialized with zeros by the network. None of the kernel regularizer, activity regularizer, bias regularizer, kernel constraint, and bias constraint is applied to any of the two dense layers. In the second Dense layer, there is an equal number of nodes as to the number of classes. The activation function used for this Dense layer is Softmax, which converts the output vector to categorical probabilities.

*4.4. Learning Rate Scheduler*

The training procedure incorporates a learning rate scheduler for better and efficient learning of image features (generated in distributed CNN). The scheduler used for training is 'cosine-annealing'. Generally, such learning rate schedulers with restarting mechanism are also known as the stochastic gradient descent with warm restarts (SGDR) [40]. But the training mechanism is using the restarting mechanism with the RMSProp optimizer. This restart technique frees the optimization from local minima over the optimization space at any time during the training. 'cosine annealing warm restart' consists of two parts. The first one is the 'cosine function', which acts as the learning rate annealing function. The second part is the 'warm-restarts', which makes the learning rate scheduler restart again from the initial point. The purpose of using such a scheduler is to maximize the probability of converging to the global minimum cost location and also to minimize the probability of being stuck at a local minimum cost point. For the current purpose, the training scheme uses an initial learning rate of 0.00001, which is the maximum learning rate. Besides, the total number of cycles within the range of total epochs is 10. As a result, the minimum learning rate achieved was only 0.00000006155. Within the $i^{th}$ run, the learning rate is decayed using the following function specified in [40].

$$\eta_t = \eta_{min}^i + \frac{1}{2}\left(\eta_{max}^i - \eta_{min}^i\right)\left(1 + cos\left(\frac{T_{cur}}{T_i}\pi\right)\right) \qquad (18)$$

where, $\eta_{max}^i$ and $\eta_{min}^i$ are ranges for the learning rate, $T_{cur}$ denotes the number of epochs that have passed since the last restart, $T_i$ denotes the periodicity of the restart (i.e. after $T_i$ epochs, the learning rate is restarted again from the initial rate). Fig. 11 represents the schematic diagram of the 'cosine-annealing' scheduler.

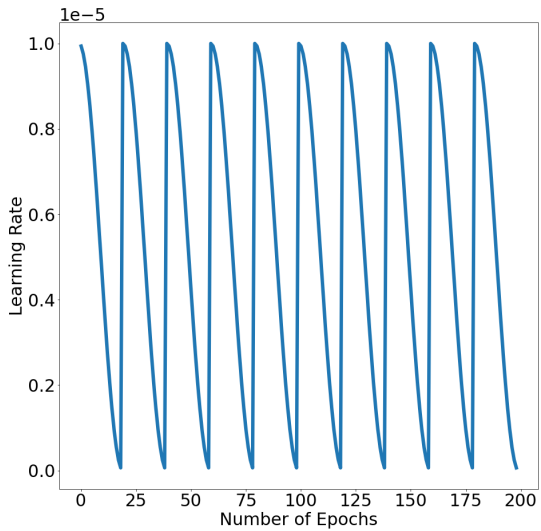

Figure 11: Cosine Annealing learning rate scheduler used for training. The maximum learning rate is 0.00001, number of cycles within the total range of epochs is 10.

*4.5. Training Parameters*

The optimizer for the neural network was RMSProp [41]. The discounting factor for the history/'coming gradient' for RMSProp was selected to be 0.9. The gradients are normalized by the un-centered second moment. The optimizer is implemented in such a way that the variables and the respective accumulators (i.e. gradient moving average, momentum, square gradient moving average, etc.) are updated even when the gradient becomes zero. The loss function used for penalizing the weight update during training was categorical cross-entropy, otherwise known as softmax loss (i.e. softmax activation and cross-entropy loss). The loss function is formulated as follows [42]-

$$\text{Categorical cross-entropy} = -log\left(\frac{e^{P_p}}{\sum_{i=1}^{M} e^{P_i}}\right) \quad (19)$$

here, M=total number of classes, $P_p$ = CNN score for the positive classes only. Chollet et al. [43] reports the implementation of the optimizer, initializer, loss functions, and other training parameters. The model is trained with a batch size of 128 and total epochs of 200, with cosine annealing as the learning rate scheduler. The neuron biases are initialized with constant zero. The kernels in convolutional layers are initialized with a normal distribution. The learning rate is kept similar to all the layers. The model is run on a single TESLA K80 Graphics Processing Unit(GPU).

*4.6. Evaluation Metrics*

For the proposed method, a variety of evaluation metrics are reported that help to evaluate the proposed method efficiently. In this subsection, a brief overview of those metrics is presented.

*4.6.1. TP, TN, FP, FN*

Four special metrics are calculated based on the predicted results. 'True-Positive (TP)' metric denotes the number of correctly classified records that belong to a specific class. 'True-Negative (TN)' metric denotes the number of correctly classified records that do not belong to a specific class. 'False-Positive (FP)' denotes the number of incorrectly classified records that do not belong to a specific class. Finally, 'False-Negative (FN)' denotes the number of incorrectly classified records that belong to a specific class. The definitions can be found in more domain-specific manner in [44]. Usually, FNR is important for biological disease classification. The False-Negative-Rate (FNR) (otherwise known as 'miss-rate') is expressed as follows-

$$FNR = \frac{FN}{FN+TP} \qquad (20)$$

*4.6.2. Categorical Accuracy*

It denotes how often the predicted class matches the true class. The proposed method in this paper has been applied on a database with 10 classes. Besides, this paper also reports the class-specific accuracy that helps to explain the class-specific performance of the models. The following equation formulates the overall accuracy in terms of TP, TN, FP, FN-

$$Accuracy = \frac{TP+TN}{N} \qquad (21)$$

here, N=total number of data.

*4.6.3. Precision or Positive Predictive Value (PPV)*

'Precision' denotes the proportion of correctly classified target birds and the total number of target and non-target birds. The following equation defines PPV-

$$Precision = \frac{TP}{TP+FP} \qquad (22)$$

*4.6.4. specificity, selectivity or true negative rate (TNR)*

'Specificity' denotes the proportion of correctly classified records that do not belong to a specific class. This can be formulated as follows-

$$Specificity = \frac{TN}{TN+FP} \qquad (23)$$

*4.6.5. F1 Score and F2 Score*

The $F_\beta$ score combines the precision and recall. If $\beta = 1$, this is known as $F_1$ score and if $\beta = 2$, it is known as $F_2$ score.. The formula is expressed as -

$$F_\beta = (1 + \beta^2) \frac{Precision.Recall}{\beta^2 Precision + Recall} \tag{24}$$

$F_1$ score provides equal importance to precision and recall. $F_2$ score provides weights recall higher than precision.

*4.6.6. Area under the curve (AUC)*

AUC measures the area under the Receiver Operating Characteristic curve (ROC curve). It provides an aggregate performance of the model. AUC does not depend on the classification threshold. However, it is also a scale-invariant parameter. The higher the value of AUC ($0 < AUC < 1$), the better the performance of the model.

**5. Results and Discussion**

This section presents the detailed result of the study. The proposed method is validated on a combination of different sets of features.

*5.1. Overall Performance*

Table 1 lists the five different feature sets over which experiment is conducted. In table 2, the average evaluation metrics for each of the feature sets are depicted.

All of the feature sets provide comparable performance to each other. The analysis based on short term feature windows, two-stage windowing, and distributed CNN-LSTM provide a similar kind of result irrespective of the feature set. However, the precision, recall, AUC, and specificity is very high for all of the classes and all feature sets. It is evident from table 2 that feature set 3 (combination of 13 MFCC features and Chroma Vectors) provides the best accuracy (90.45%), precision (90.45%), recall (90.45%), F1 score (90.45%), F2 score (90.45%), specificity (98.94%), AUC (94.09%), and the lowest FNR (9.55%).

Table 1: Different sets of feature sets used as training data.

| Feature Sets | Specific Features |
|---|---|
| Feature Set 1 | 13 MFCC |
| Feature Set 2 | 12 Chroma Vectors |
| Feature Set 3 | 13 MFCC + 12 Chroma Vectors |
| Feature Set 4 | Spectral Features (Spectral Centroid, Spectral Entropy, Spectral Flux, Spectral Roll-off, Spectral Spread) |
| Feature Set 5 | All Features (13 MFCC, 12 Chroma Vectors, Spectral features, Chroma Deviation, Energy, Entropy of Energy, Zero-Crossing Rate) |

Table 2: Classification performance for different feature sets as training data.

| Feature Set | Accuracy (%) | Specificity (%) | F1 score (%) | FNR (%) | AUC (%) | Precision (%) |
|---|---|---|---|---|---|---|
| Feature Set 1 | 89.94 | 98.88 | 89.94 | 10.06 | 93.91 | 89.94 |
| Feature Set 2 | 88.07 | 98.67 | 88.07 | 11.93 | 92.83 | 88.07 |
| Feature Set 3 | **90.45** | **98.94** | **90.45** | **9.55** | **94.09** | **90.45** |
| Feature Set 4 | 90.09 | 98.90 | 90.09 | 9.91 | 93.92 | 90.09 |
| Feature Set 5 | 89.98 | 98.89 | 89.98 | 10.02 | 93.85 | 89.98 |

The second best feature set turns out to be the spectral features (feature set 4). The accuracy, specificity, and AUC for this set are 90,09%, 98.90%, and 93.92%, respectively. Next, a standalone feature set consisting of only 13 MFCC feature set provides an accuracy of 89.94%, the specificity of 98.90%, and AUC of 93.91%. Although MFCC and chroma vectors provide comparable performances with each other, their combination achieves the best performance.

*5.2. Class Specific Performance*

Fig. 12 represents the class-specific evaluation metrics for different sets of features. Fig . 12(a) illustrates the class specific performance of the model. All of the feature sets show almost similar performance in terms of accuracy.Figure 12(b) shows the false-negative rate for each class. Although, for some specific class (e.g., Pygmy Nuthatch) the false negativity rate increases more than any other class, there is performance difference among different feature sets. On an average, feature set 3 achieves the lowest false negativity. Similarly, the F1 score in 12(c) and the precision in 12(d) and the AUC in 12(f) depict the overall superior performance of feature set 3. All the feature sets show almost similar performance in terms of specificity 12(e).

*5.3. Performance Comparison*

Several studies such as [3],[45],[46],[47],[48],[49],[50],[51] represent the performance bird call/song classification task from the perspective of traditional signal processing or machine learning or both. However, any comparison among them is quite ambiguous since different methods use different number of classes and records. Also, there are issues of class imbalance, data quality and evaluation metrics. One of the main purpose of this paper is to focus on acoustic features and decide which set of feature best captures the inherent identity of any bird song/call. It seems a combination of MFCC and chroma vectors could be the best choice of acoustic features for any deep-learning-based bird call/song classification task.

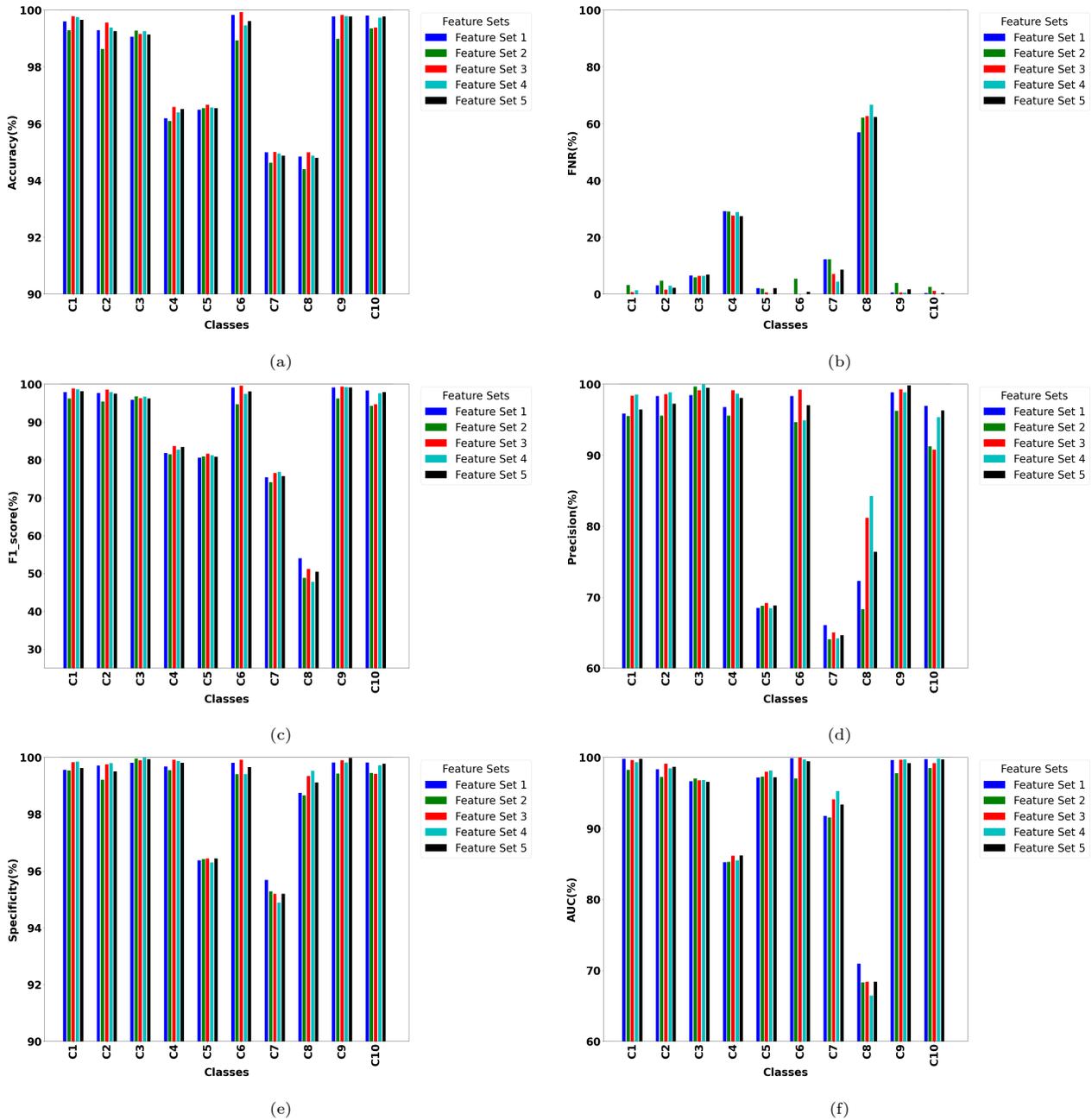

Figure 12: Class specific performance of the proposed method.(a) Accuracy, (b) False Negative Rate, (c) F1 score, (d) Precision, (e) Specificity, (f) AUC. The accuracy and specificity show a similar performance in various class irrespective of the feature set. However, FNR, F1 score, precision, and AUC show a overall superior performance of feature set 3 to other feature sets. Here, C1,C2,...,C10 represent the class numbers.

## 6. Conclusion

The proposed method involves two-stage windowing that generates a two-dimensional feature image for the distributed CNN-LSTM network. The network classifies the bird species based on the signal of their call/songs. One of the purposes of this study is to select the acoustic feature set, which provides better results than other features. The paper has an analysis of 34 acoustic features. Among them, a combination of 13 MFCC features and 12 Chroma Vector features achieves the best results in terms of accuracy, precision, recall, $F_\beta$ scores, specificity, AUC, and FNR. The second purpose of the analysis is to study the effect of the adaptive thresholding technique for voice activity detection in this problem. The third purpose is to find out the performance of a distributed CNN-LSTM network. The CNN features are suitable for image feature extraction, and LSTM is suitable for temporal feature extraction. The task of bird species classification from bird acoustics has always been an interesting research area that requires deeper analysis of bird acoustic features. In the recently reported studies, researchers always selected a certain feature that is used for bird species classification through a simple neural network, whereas other features or their combination might have significant and insightful results. This paper tries to compare the results by thoroughly investigating and analyzing different combinations of 34 bird acoustic features. It is helpful for better identification and classification of various bird species in an environmental setting. The future study can extend the same analysis to a large-scale bird species classification with data from different sources. Besides, the same technique could be applied to any acoustic applications other than bird calls/songs.

## 7. Acknowledgement

Declarations of interest: none. This research did not receive any specific grant from funding agencies in the public, commercial, or not-for-profit sectors.

**References**

[1] R. H. Zottesso, Y. M. Costa, D. Bertolini, L. E. Oliveira, Bird species identification using spectrogram and dissimilarity approach, Ecological Informatics 48 (2018) 187–197.

[2] S. Kahl, T. Wilhelm-Stein, H. Hussein, H. Klinck, D. Kowerko, M. Ritter, M. Eibl, Large-scale bird sound classification using convolutional neural networks., in: CLEF (Working Notes), 2017.

[3] C.-Y. Koh, J.-Y. Chang, C.-L. Tai, D.-Y. Huang, H.-H. Hsieh, Y.-W. Liu, Bird sound classification using convolutional neural networks., in: CLEF (Working Notes), 2019.

[4] K. He, X. Zhang, S. Ren, J. Sun, Deep residual learning for image recognition, in: Proceedings of the IEEE conference on computer vision and pattern recognition, 2016, pp. 770–778.

[5] C. Szegedy, V. Vanhoucke, S. Ioffe, J. Shlens, Z. Wojna, Rethinking the inception architecture for computer vision, in: Proceedings of the IEEE conference on computer vision and pattern recognition, 2016, pp. 2818–2826.

[6] A. Incze, H.-B. Jancsó, Z. Szilágyi, A. Farkas, C. Sulyok, Bird sound recognition using a convolutional neural network, in: 2018 IEEE 16th International Symposium on Intelligent Systems and Informatics (SISY), IEEE, 2018, pp. 000295–000300.

[7] A. G. Howard, M. Zhu, B. Chen, D. Kalenichenko, W. Wang, T. Weyand, M. Andreetto, H. Adam, Mobilenets: Efficient convolutional neural networks for mobile vision applications, arXiv preprint arXiv:1704.04861 (2017).

[8] J. Xie, M. Zhu, Handcrafted features and late fusion with deep learning for bird sound classification, Ecological Informatics 52 (2019) 74–81.

[9] N. Bold, C. Zhang, T. Akashi, Cross-domain deep feature combination for bird species classification with audio-visual data, IEICE TRANSACTIONS on Information and Systems 102 (10) (2019) 2033–2042.

[10] M. Lasseck, Acoustic bird detection with deep convolutional neural networks, in: Proceedings of the Detection and Classification of Acoustic Scenes and Events 2018 Workshop (DCASE2018), 2018, pp. 143–147.

[11] F. Briggs, R. Raich, X. Z. Fern, Audio classification of bird species: A statistical manifold approach, in: 2009 Ninth IEEE International Conference on Data Mining, IEEE, 2009, pp. 51–60.

[12] M. Raghuram, N. R. Chavan, R. Belur, S. G. Koolagudi, Bird classification based on their sound patterns, International journal of speech technology 19 (4) (2016) 791–804.

[13] J. Schlüter, Bird identification from timestamped, geotagged audio recordings., in: CLEF (Working Notes), 2018.

[14] S. HEUER, P. TAFO, H. HOLZMANN, S. DAHLKE, New aspects in birdsong recognition utilizing the gabor transform, in: Proceedings of the 23rd International Congress on Acoustics. Aachen, 2019.

[15] G. Peeters, A large set of audio features for sound description (similarity and classification) in the cuidado project, CUIDADO IST Project Report 54 (0) (2004) 1–25.

[16] J.-l. Shen, J.-w. Hung, L.-s. Lee, Robust entropy-based endpoint detection for speech recognition in noisy environments, in: Fifth international conference on spoken language processing, 1998.

[17] E. Scheirer, M. Slaney, Construction and evaluation of a robust multifeature speech/music discriminator, in: 1997 IEEE international conference on acoustics, speech, and signal processing, Vol. 2, IEEE, 1997, pp. 1331–1334.

[18] F. E. Beddard, The structure and classification of birds, Longmans, Green, and Company, 1898.

[19] L. Rabiner, R. Schafer, Theory and applications of digital speech processing, Prentice Hall Press, 2010.

[20] P. Mermelstein, Distance measures for speech recognition, psychological and instrumental, Pattern recognition and artificial intelligence 116 (1976) 374–388.

[21] B. Logan, et al., Mel frequency cepstral coefficients for music modeling., in: Ismir, Vol. 270, Citeseer, 2000, pp. 1–11.

[22] M. R. Hasan, M. Jamil, M. Rahman, et al., Speaker identification using mel frequency cepstral coefficients, variations 1 (4) (2004).

[23] D. Chazan, R. Hoory, G. Cohen, M. Zibulski, Speech reconstruction from mel frequency cepstral coefficients and pitch frequency, in: 2000 IEEE International Conference on Acoustics, Speech, and Signal Processing. Proceedings (Cat. No. 00CH37100), Vol. 3, IEEE, 2000, pp. 1299–1302.

[24] L. Muda, M. Begam, I. Elamvazuthi, Voice recognition algorithms using mel frequency cepstral coefficient (mfcc) and dynamic time warping (dtw) techniques, arXiv preprint arXiv:1003.4083 (2010).

[25] M. Müller, Fundamentals of music processing: Audio, analysis, algorithms, applications, Springer, 2015.

[26] T. Cho, J. P. Bello, On the relative importance of individual components of chord recognition systems, IEEE/ACM Transactions on Audio, Speech, and Language Processing 22 (2) (2013) 477–492.

[27] M. Mauch, S. Dixon, Simultaneous estimation of chords and musical context from audio, IEEE Transactions on Audio, Speech, and Language Processing 18 (6) (2009) 1280–1289.

[28] D. Ellis, Chroma feature analysis and synthesis, Resources of Laboratory for the Recognition and Organization of Speech and Audio-LabROSA (2007).

[29] N. Jiang, P. Grosche, V. Konz, M. Müller, Analyzing chroma feature types for automated chord recognition, in: Audio Engineering Society Conference: 42nd International Conference: Semantic Audio, Audio Engineering Society, 2011.

[30] M. A. Bartsch, G. H. Wakefield, Audio thumbnailing of popular music using chroma-based representations, IEEE Transactions on multimedia 7 (1) (2005) 96–104.

[31] M. Müller, Music synchronization, in: Fundamentals of music processing, Springer, 2015, Ch. 3, pp. 115–166.

[32] Center for conservation bioacoustics, `https://www.birds.cornell.edu/ccb/`, accessed: January 1, 2022(2020).

[33] The cornell lab of ornithology, `https://www.birds.cornell.edu/home/`, accessed: January 1, 2022(2020).

[34] B. L. Sullivan, C. L. Wood, M. J. Iliff, R. E. Bonney, D. Fink, S. Kelling, ebird: A citizen-based bird observation network in the biological sciences, Biological conservation 142 (10) (2009) 2282–2292.

[35] K. Sakhnov, E. Verteletskaya, B. Simak, Approach for energy-based voice detector with adaptive scaling factor., IAENG International Journal of Computer Science 36 (4) (2009).

[36] X. Glorot, Y. Bengio, Understanding the difficulty of training deep feedforward neural networks, in: Proceedings of the thirteenth international conference on artificial intelligence and statistics, 2010, pp. 249–256.

[37] B. Xu, N. Wang, T. Chen, M. Li, Empirical evaluation of rectified activations in convolutional network, arXiv preprint arXiv:1505.00853 (2015).

[38] S. Hochreiter, J. Schmidhuber, Long short-term memory, Neural computation 9 (8) (1997) 1735–1780.

[39] R. Jozefowicz, W. Zaremba, I. Sutskever, An empirical exploration of recurrent network architectures, in: International conference on machine learning, 2015, pp. 2342–2350.

[40] I. Loshchilov, F. Hutter, Sgdr: Stochastic gradient descent with warm restarts, arXiv preprint arXiv:1608.03983 (2016).

[41] G. Hinton, N. Srivastava, K. Swersky, Neural networks for machine learning lecture 6a overview of mini-batch gradient descent, Cited on 14 (8) (2012).

[42] Z. Zhang, M. Sabuncu, Generalized cross entropy loss for training deep neural networks with noisy labels, in: Advances in neural information processing systems, 2018, pp. 8778–8788.

[43] F. Chollet, et al., Keras, accessed: January 1, 2022(2015).
URL https://github.com/fchollet/keras

[44] K. M. Ting, Confusion Matrix, Springer US, Boston, MA, 2017, pp. 260–260. doi:10.1007/978-1-4899-7687-1_50.
URL https://doi.org/10.1007/978-1-4899-7687-1_50

[45] D. B. Efremova, M. Sankupellay, D. A. Konovalov, Data-efficient classification of birdcall through convolutional neural networks transfer learning, in: 2019 Digital Image Computing: Techniques and Applications (DICTA), IEEE, 2019, pp. 1–8.

[46] M. Sankupellay, D. Konovalov, Bird call recognition using deep convolutional neural network, resnet-50, in: Proceedings of ACOUSTICS, Vol. 7, 2018.

[47] W. Chu, D. T. Blumstein, Noise robust bird song detection using syllable pattern-based hidden markov models, in: 2011 IEEE International Conference on Acoustics, Speech and Signal Processing (ICASSP), IEEE, 2011, pp. 345–348.

[48] H. Goëau, S. Kahl, H. Glotin, R. Planqué, W.-P. Vellinga, A. Joly, Overview of birdclef 2018: monospecies vs. soundscape bird identification, in: CLEF: Conference and Labs of the Evaluation Forum, no. 2125, 2018.

[49] S. Kahl, T. Wilhelm-Stein, H. Klinck, D. Kowerko, M. Eibl, A baseline for large-scale bird species identification in field recordings., in: CLEF (Working Notes), 2018.

[50] S. Kahl, F.-R. Stöter, H. Goëau, H. Glotin, R. Planque, W.-P. Vellinga, A. Joly, Overview of birdclef 2019: large-scale bird recognition in soundscapes, in: Working Notes of CLEF 2019-Conference and Labs of the Evaluation Forum, no. 2380, CEUR, 2019, pp. 1–9.

[51] D. Stowell, M. D. Plumbley, Automatic large-scale classification of bird sounds is strongly improved by unsupervised feature learning, PeerJ 2 (2014) e488.